\documentclass[pre,twocolumn,showpacs]{revtex4}
\usepackage{graphicx}% Include figure files
\usepackage{dcolumn}% Align table columns on decimal point
\usepackage{bm}% bold math
\usepackage{amsmath, amsthm, amssymb}
\usepackage{psfig}
\usepackage{epsfig}
\begin{document}
\def\kket{\rangle \mskip -3mu \rangle}
\def\bbra{\langle \mskip -3mu \langle}

\def\ket{\rangle}
\def\bra{\langle}

\def\pard{\partial}

\def\sinh{{\rm sinh}}
\def\sgn{{\rm sgn}}

%Characters
\def\alp{\alpha}
\def\del{\delta}
\def\Del{\Delta}
\def\eps{\epsilon}
\def\gam{\gamma}
\def\sig{\sigma}
\def\kap{\kappa}
\def\lam{\lambda}
\def\ome{\omega}
\def\Ome{\Omega}

\def\th{\theta}
\def\vphi{\varphi}

\def\Gam{\Gamma}
\def\Ome{\Omega}

\def\kav{{\bar k}}
\def\vb{{\bar v}}

\def\abf{{\bf a}}
\def\cbf{{\bf c}}
\def\dbf{{\bf d}}
\def\gbf{{\bf g}}
\def\kbf{{\bf k}}
\def\lbf{{\bf l}}
\def\nbf{{\bf n}}
\def\pbf{{\bf p}}
\def\qbf{{\bf q}}
\def\rbf{{\bf r}}
\def\ubf{{\bf u}}
\def\vbf{{\bf v}}
\def\xbf{{\bf x}}
\def\Cbf{{\bf C}}
\def\Dbf{{\bf D}}
\def\Kbf{{\bf K}}
\def\Pbf{{\bf P}}
\def\Qbf{{\bf Q}}

\def\omet{{\tilde \ome}}
\def\gammat{{\tilde \gamma}}
\def\Ft{{\tilde F}}
\def\ut{{\tilde u}}
\def\bt{{\tilde b}}
\def\vt{{\tilde v}}
\def\xt{{\tilde x}}

\def\ph{{\hat p}}

\def\vt{{\tilde v}}
\def\wt{{\tilde w}}
\def\phit{{\tilde \phi}}
\def\rhot{{\tilde \rho}}
\def\Ft{ {\tilde F}}

\def\Cb{{\bar C}}
\def\Nb{{\bar N}}
\def\Ab{{\bar A}}
\def\Db{{\bar D}}
\def\etab{{\bar \eta}}
\def\gb{{\bar g}}
\def\nb{{\bar n}}
\def\bb{{\bar b}}
\def\Pib{{\bar \Pi}}
\def\rhob{{\bar \rho}}?\def\phib{{\bar \phi}}
\def\psib{{\bar \psi}}
\def\omeb{{\bar \ome}}

\def\Sh{{\hat S}}
\def\Wh{{\hat W}}

\def\SS{I}
\def\psiw{{\xi}}
\def\tI{{g}}

\def\Ep#1{Eq.\ (\ref{#1})}
\def\Eqs#1{Eqs.\ (\ref{#1})}
\def\EQN#1{\label{#1}}

\newcommand{\beqa}{\begin{eqnarray}}
\newcommand{\eeqa}{\end{eqnarray}}

%\preprint{APS/123-QED}

\title{A new way of defining unstable states}% Force line breaks with \\

\author{Sungyun Kim}
 \email{ksyun@mpipks-dresden.mpg.de}
 \affiliation{Max-Planck Institute for Physics of Complex Systems, N\"{o}thnizter Str 38, 01187
  Dresden, Germany }%Lines break automatically or can be forced with \\
\author{Gonzalo Ordonez}%
 \email{gonzalo@physics.utexas.edu}
\affiliation{%
Center for Studies in Statistical Mechanics and Complex Systems,
The University of Texas at Austin, Austin, Texas 78712 USA}
\date{\today}
\begin{abstract}
We define a new unstable state in the Friedrichs model of a two-level atom. This
unstable state is a complex eigenstate of the time evolution
operator $\exp(-iHt)$ with a restricted test function space, which is obtained from causality conditions. The
unstable state
 shows exact exponential decay for $t\ge0$. Its emitted field is confined inside the future light-cone.
In this way the
long-standing problem of exponential catastrophe is removed.  This is an example of quantum mechanics outside Hilbert space,
 which consists of generalized eigenstates in a distribution space, and
 a dual (test function) space.
\end{abstract}
\maketitle

%\section{\label{sec:level1}First-level heading:\protect\\ The line
%break was forced \lowercase{via} \textbackslash\textbackslash}

\section{Introduction}
 The problem of defining unstable states has a long and
controversial history. Although unstable particles abound in
nature, usual quantum states with real energies in Hilbert
space describe only stable particles. Thus the question arises
whether we can construct an unstable state that represents an
unstable particle.

 A number of people have studied this problem. Gamow first
 introduced complex energies to model unstable particles with
 exponential decay \cite{Gamow}. Nakanishi \cite{Nakanishi} introduced complex distributions to
 define a complex eigenstate of the Hamiltonian in Lee's model \cite{Lee}. The real part of the eigenvalue gave
 the particle's mass, and the imaginary part gave the lifetime.
 In this way, a state with complex eigenvalue represented the
 unstable state.
 Sudarshan, Chiu and Gorini
  \cite{Sudarshan} constructed complex eigenstates  using contour
  deformation in the complex plane. Bohm and Gadella \cite{Bohm} constructed complex
  eigenvectors using poles of the $S$ matrix and Hardy class test functions (see also \cite{Tasaki}).
  Prigogine and collaborators
   studied extensively the properties of complex spectral representations in
  the Friedrichs model \cite{PPT}, and defined unstable
  states in Liouville space (see \cite{OPP} and references therein).
   Still, the exponential growth of the field component outside the
   light cone (also called the exponential catastrophe) remained
    as a problem \cite{symmetry1,symmetry2}.

  In this article we show another way of constructing an unstable state without
  exponential catastrophe in the Friedrichs model. This is done by separating the
  pole contribution using a suitable integration contour and test function space.
  The state we construct becomes a complex eigenstate of the
  the time evolution operator $e^{-iHt}$ within
  a suitable test function space.

 The paper is organized as follows. In section
 II we explain the Friedrichs model. In section
 III we review a previous approaches based on
 contour deformation \cite{Sudarshan,Tasaki} (in Appendix A we review other approach
 based on a ``rigged'' Hilbert space with Hardy-class test functions \cite{Bohm}).
 We point out difficulties of these approaches in describing unstable
 states. In section IV, we propose another way of taking the
 complex pole and show that this method eliminates the exponential growth.
 In section V, we conclude our result and discuss the extension of
 quantum mechanics outside the Hilbert space.

 \section{Model}

  We consider the Friedrichs model in one dimension \cite{Fano, Friedrichs}.
  This is a simplified version of the Lee model of unstable particle in
  the one-particle sector \cite{Lee}. It is also a model of a two-level atom
   interacting with the electromagnetic field
  in the dipole and rotating wave approximations \cite{cohen}. Hereafter we focus on the atom-photon interpretation of the model.
   The Hamiltonian is given by

 \beqa
& &H_F = \ome_1 |1\ket\bra 1| + \int_{-\infty}^{\infty} dk \;
\ome_k |k\ket \bra k|
 \nonumber \\& &+ \lam
 \int_{-\infty}^{\infty} dk \; \vb_k (|1\ket \bra k| + |k\ket \bra 1|) \EQN{1}
 \eeqa
where we put $c=\hbar =1$. The state $|1\ket$ represents the bare
atom in its excited level with no field present, while the
state $|k\ket$ represents a bare field mode (``photon'') of
momentum $k$ together with the atom in its ground state. The excited state is analogous to an unstable particle state, while the photon is analogous to decay products.

The energy of the ground state is chosen to be zero; $\ome_1$  is
the bare energy of the excited level and $\ome_k\equiv |k|$ is the
photon energy. $\lam$ is a small dimensionless coupling constant
($\lam \ll 1$). We shall consider a specific form of the
interaction potential
\begin{eqnarray}
  \vb_k  =  \frac{ \ome_k^{1/2}}{ 1 +(\ome_k/M)^2}.
 \EQN{4}
\end{eqnarray}
The constant $M^{-1}$ determines the range of the interaction and
gives an ultraviolet cutoff. Other forms of potential (form factors) may be treated in a similar way, the only condition being that they are exponentially bounded at infinity, as we will discuss later on.

 From the dispersion relation $\ome_k = |k|$, the free-Hamiltonian
 eigenstates $|k\ket$ and $|-k\ket$ have the same eigenvalue $\ome_k$. We
 remove this degeneracy by rewriting the Hamiltonian
  \beqa
   & &H_F = \ome_1 |1\ket \bra 1| + \int_{0}^{\infty} dk\; \ome_k
   \left(|S_{k}\ket \bra S_{k}| + |A_{k}\ket \bra A_{k}| \right) \nonumber \\
    & & + \int_{0}^{\infty} dk\;\sqrt{2}\lam \vb_{k}
   (|1\ket \bra S_{k}|+|S_{k}\ket\bra 1|)
   \EQN{5-1}
 \eeqa
 where
 \beqa
  |S_{k}\ket  \equiv \frac{1}{\sqrt{2}} (|k\ket + |-k\ket),\;\;\;
  |A_{k}\ket  \equiv \frac{1}{\sqrt{2}} (|k\ket - |-k\ket). \EQN{5-2}
  \eeqa
 From \Ep{5-1} we see that the discrete eigenstate $|1\ket$ only
 interacts with the symmetric field eigenstate $|S_{k}\ket$.
 Also in this form the Hamiltonian is expressed with energy
 eigenstates, not the $k$ eigenstates.
 The anti-symmetric field component acts like a free field and can
 be treated separately. From now on, we concentrate on only the
 discrete atom state and the symmetric field states of the Hamiltonian.
 Changing integration variable $k$ to $\ome$ and rewriting
 \beqa
  |\ome\ket \equiv |S_{k}\ket,\;\;\; v_{\ome} \equiv
  \sqrt{2}\vb_{\ome_k}, \EQN{2-6}
  \eeqa
 we get the atom-field interaction Hamiltonian
  \beqa
   & & H \equiv \ome_1 |1\ket \bra 1| + \int_{0}^{\infty} d\ome\; \ome
   |\ome\ket \bra \ome| \nonumber\\
    & &+ \int_{0}^{\infty} d\ome\;\lam v_{\ome}
   (|1\ket \bra {\ome}|+|{\ome}\ket\bra 1|). \EQN{2-7}
   \eeqa

 This Hamiltonian has an exact diagonalized form. When the
 atom eigenfrequency $\ome_1$ is outside the field spectrum
 ( $\ome_1 <0$) , we call this stable case. In this case the Hamiltonian is diagonalized as
 \beqa
 H_s = \omeb_1 |\phib_1 \ket \bra \phib_1| + \int_{0}^{\infty} d\ome \;\ome
 |\phib_{\ome}^{\pm}\ket\bra\phib_{\ome}^{\pm}| \EQN{5}
 \eeqa
 where $|\phib_1 \ket$ and $|\phib_{\ome}\ket $ are given by
 \beqa
 |\phib_1\ket = \Nb_1^{1/2}\left( |1\ket + \int_{0}^{\infty} d\ome
 \frac{\lam v_{\ome} |\ome\ket}{\omeb_1 - \ome} \right), \EQN{6}
 \eeqa
 \beqa
 |\phib_{\ome}^{\pm} \ket = |\ome\ket + \frac{\lam v_{\ome}}{\eta^{\pm}(\ome)}|1\ket
 +\frac{\lam v_\ome}{\eta^{\pm}(\ome)} \int_{0}^{\infty} d\ome'
 \frac{\lam v_{\ome'} |\ome'\ket}{\ome-\ome' \pm i\eps} \EQN{7}
 \eeqa
   with
 \beqa
 \Nb_1 \equiv \left( 1+ \int_{0}^{\infty} d\ome \frac{ \lam^2
 v_\ome^2}{(\omeb_1-\ome)^2} \right)^{-1}, \EQN{8}
 \eeqa
 \beqa
 \eta^{\pm}(z) \equiv z-\ome_1 -\int_{0}^{\infty} d\ome
 \frac{\lam^2 v_\ome^2}{z^{\pm}-\ome}. \EQN{9}
 \eeqa
 We can choose $+$ branch or $-$ branch for the diagonalized solution.
  These branches
 correspond to  outgoing   and incoming waves, respectively. In \Ep{9},
 $1/(z^{\pm}-\ome)$ means that $z$ is analytically
 continued from above ($+$) or below ($-$). For real z, it can be
 understood as
 \beqa
  \frac{1}{z^{\pm}-\ome }= \lim_{\eps \rightarrow 0^+}
  \frac{1}{z\pm i\eps-\ome } \EQN{9-1}
  \eeqa
 For the stable case ($\ome_1<0$),
 the diagonalized Hamiltonian has a renormalized atom state
 $|\phib_1\ket$ with renormalized atom frequency $\omeb_1 <0$ satisfying the relation
 \beqa
  \omeb_1-\ome_1-\int_0^{\infty} d\ome\; \frac{\lam^2
  v_{\ome}^2}{\omeb_1-\ome} =0 \EQN{9-2}
 \eeqa

 When the atom eigenvalue $\ome_1$ is inside the continuum,
 the situation changes. For
 \beqa
 \ome_1 > \int_{0}^{\infty} d\ome
 \frac{\lam^2 v_\ome^2 }{\ome}, \EQN{10-1}
 \eeqa
  the equation $\eta^{\pm}(z) =0$ does
 not have a real solution. Unlike stable case, we cannot maintain
 the renormalized atom state with real eigenvalue.

 One diagonalized solution for this case is due to Friedrichs \cite{Friedrichs}, and has
the form
 \beqa
 & &H = \int_{0}^{\infty} d\ome \; \ome |F_\ome^{\pm}\ket \bra F_\ome^{\pm}|, \EQN{11} \\
 & &|F_\ome^{\pm} \ket = |\ome\ket + \frac{\lam v_\ome }{\eta^{\pm}(\ome)}
 \left( |1\ket +  \int_{0}^{\infty} d\ome' \frac{\lam v_{\ome'}}{\ome -
 \ome' \pm i \eps}|{\ome'}\ket \right). \EQN{12}\nonumber \\
 \eeqa

 The eigenstates satisfy the eigenvalue equation as well as
  the orthonormality and completeness relations \cite{PPT}
  \beqa
   H|F_\ome^{\pm}\ket &=& \ome |F_\ome^{\pm}\ket, \EQN{14} \\
   \bra F_\ome^\pm |F_{\ome'}^\pm\ket &=& \delta (\ome-\ome')\nonumber\\
   \int_{0}^{\infty}d\ome\;
   |F_\ome^\pm\ket \bra F_\ome^\pm| &=& |1\ket\bra 1|+
    \int_{0}^{\infty}d\ome\; |{\ome}\ket\bra {\ome}| . \EQN{15}
   \eeqa

  Note that this solution contains only field modes. The bare unstable
 atom is viewed as a superposition of the field modes.
 The difficulty is that in this view the unstable state has the memory
 of its creation.
 The decay law is not strictly exponential, and we can
 distinguish old atoms and young atoms. According to this view,
 unstable particles are also distinguished by
their creation process. Because of these complications, we
 want another definition of unstable states
 describing  indistinguishable  particles which are independent of their
 creation process. This requires strict exponential decay with no
 memory \cite{IP80}.

\begin{figure}[htb] % Imported eps example.
\begin{center}
\psfig{file=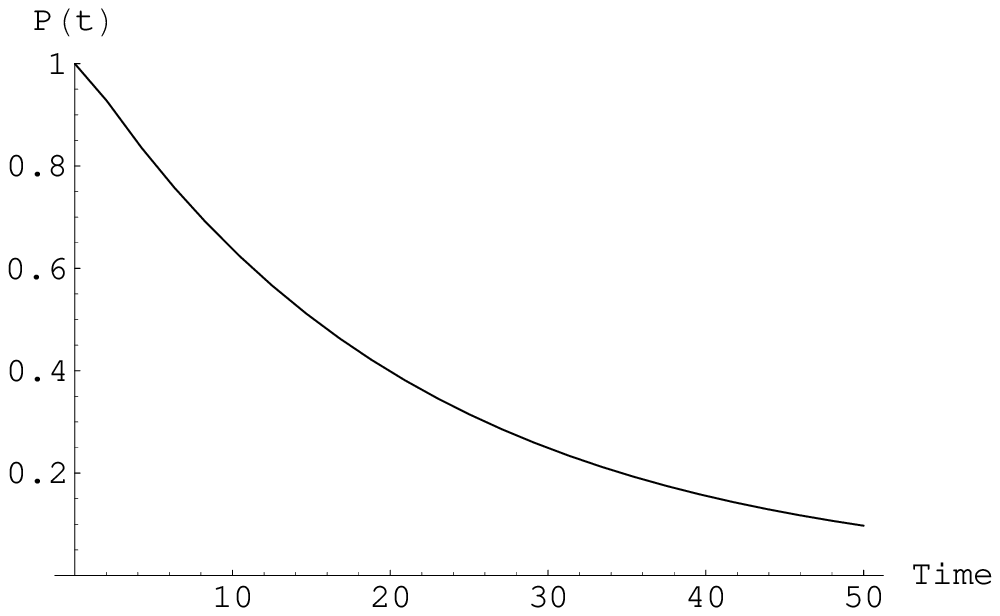, height=2.5in, width=3in}
 %unstable_fig1.ps

 \caption{The decay of survival probability $|\bra 1|e^{-iHt}|1\ket|^2$.  } \label{pdecay}
\end{center}
\end{figure}

\begin{figure}[htb] % Imported eps example.
\begin{center}
\psfig{file=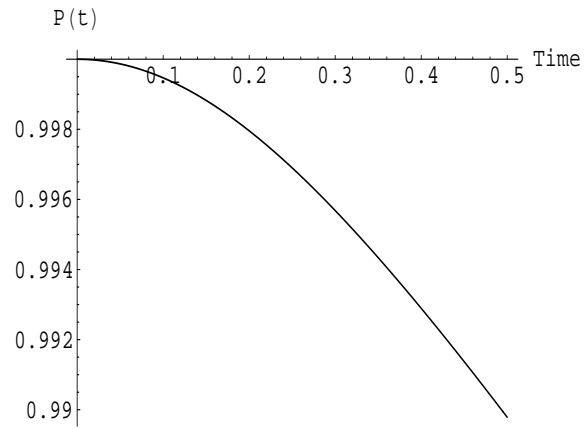, height=2.5in, width=3in}
 %unstable_fig1.ps

 \caption{The survival probability of excited atom $|\bra 1|e^{-iHt}|1\ket|^2$ near  $t=0$.
 In this short time period (Zeno time), the decay is not exponential.  } \label{zenotime}
\end{center}
\end{figure}

\begin{figure}[htb] % Imported eps example.
\begin{center}
\psfig{file=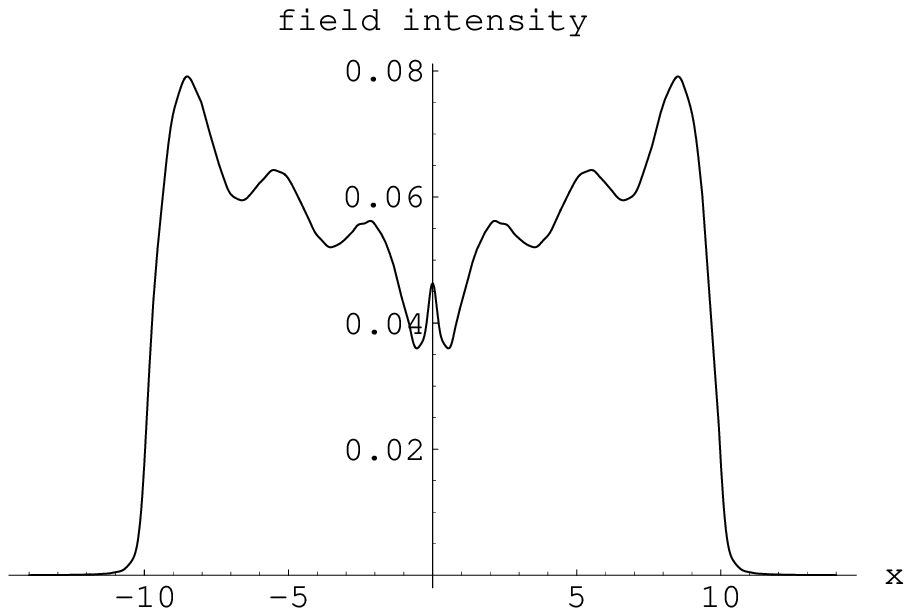, height=2.5in, width=3in}
 %unstable_fig1.ps

 \caption{The field intensity $|\bra \psi (x)|e^{-iHt}|1\ket|^2$ in space at $t=10$.}
  \label{fieldt10}
\end{center}
\end{figure}

  In figure \ref{pdecay}   we show the survival probability of the state $|1\ket$.
 As seen in figure \ref{zenotime}, the survival probability shows
  non-exponential decay around $t=0$ (Zeno effect).  Figure  \ref{fieldt10} shows
  the field generated by the initial condition of excited
  atom state and no field. We define the field bra $\bra \psi (x)|$ as
  \beqa
  \bra \psi (x) | &\equiv& \int_{-\infty}^{\infty}dk\;
  \frac{1}{\sqrt{2\ome_k}}e^{i k x} \bra k| \nonumber\\
   &=& \int_0^{\infty} d\ome\;
  \sqrt{\frac{2}{\ome}}\cos(\ome x) \bra \ome |,
 \EQN{21-x}
  \eeqa
 The
  generated field is a superposition  of field associated with the Zeno effect, exponential
  field due to spontaneous emission and dressing cloud around the atom at $x=0$ \cite{POP}.
  Note that the field disappears rapidly outside the light cone, defined by $|x|= ct$ with $c=1$.


 One way to get pure exponential decay  is to construct
 eigenstates with complex eigenvalues. This has been done
 already, but previous constructions had their own difficulties,
 for example the  exponential growth of the emitted field outside the light cone (exponential catastrophe).
 In the following sections, we review the approach based on contour deformation.

 \section{Unstable state using contour deformation }

In this section we review the construction of unstable state
through contour deformation. The construction of complex
 eigenstate through contour deformation was
 done by Sudarshan, Chiu, and Gorgini \cite{Sudarshan}. Later by
 the perturbation expansion with
  regularization rules, Petrosky, Prigogine and Tasaki \cite{PPT}
  investigated the Friedrichs model and showed that the system can
  be described as a sum of a discrete complex eigenstate plus
  continuum states. They  showed that their decomposition
  can be also derived by contour deformation. Let us follow
  their construction in the Friedrichs model.

\begin{figure}[htb] % Imported eps example.
\begin{center}
\psfig{file=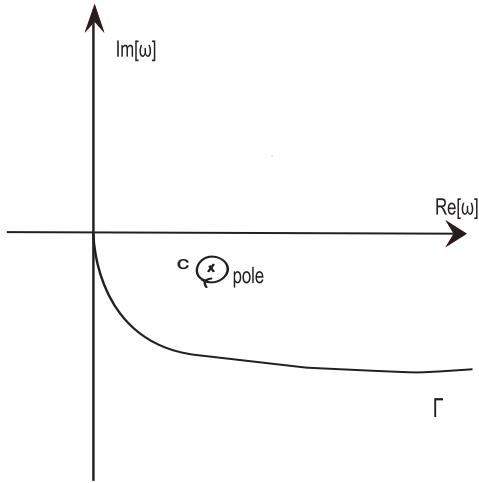, height=2.5in, width=3in}
 %unstable_fig1.ps

 \caption{ the contours $\Gamma$ and $C$ } \label{pcontour}
\end{center}
\end{figure}

   In the Friedrichs model, first we note that  $\eta^+(z)=0$ has a
    complex root $z_1 =\omet_1-i\gamma=\ome_1+ O(\lam^2)$
  for $0< \ome_1 \sim O(1)$. We can consider this as a complex
  eigenvalue which coincides with the original discrete
  eigenvalue $\ome_1$ in the limit $\lam \rightarrow 0$. By
  contour deformation, we can
  separate $z_1$ pole from $1/\eta^+(z)$ in the completeness
  relation.
 \beqa
  1 &=& \int_0^{\infty} d\ome\;|F_\ome^+\ket\bra F_\ome^+| \nonumber\\
     &=&
  \int_\Gamma d\ome\;
  |F_\ome^+\ket \bra F_\ome^+| + \int_C
 d\ome\;   |F_\ome^+\ket \bra F_\ome^+| \EQN{16}
 \eeqa
 where the contours $\Gamma$ and $C$ are shown in
 Fig~\ref{pcontour}.

 The pole part of the contour $\int_C$ can be written as
\beqa
 \int_C
 d\ome\;   |F_\ome^+\ket \bra F_\ome^+| = |\phi_1 \ket \bra
 \phit_1| \EQN{17}
 \eeqa
 where
 \beqa
 & &|\phi_1\ket =  N_1^{1/2}\left( |1\ket + \int_{0}^{\infty} d\ome
 \frac{\lam v_\ome |\ome\ket}{z_1^+ - \ome} \right), \EQN{18}
 \\
 & &\bra\phit_1| =  N_1^{1/2}\left( \bra 1 | + \int_{0}^{\infty} d\ome
 \frac{ \lam v_\ome \bra \ome |}{z_1^+ - \ome} \right),
 \EQN{19}\\
 & &N_1 \equiv \left( 1+ \int_{0}^{\infty} d\ome \frac{\lam^2
 v_\ome^2}{(z_1^+-\ome)^2} \right)^{-1}. \EQN{17-1}
  \eeqa
 This complex eigenstate $|\phi_1\ket$ satisfies the eigenvalue
 equation
  \beqa
   H|\phi_1\ket = z_1 |\phi_1\ket. \EQN{18-1}
   \eeqa
Note that $|\phi_1\ket$ cannot be in the Hilbert space since it
has a complex eigenvalue.

 It should be noted that when we act a function on $|\phi_1\ket$
 the function should not blow up on the deformed contour. We need
 suitable test functions depending on our choice of contour
 deformation. Without the test function consideration, unphysical
 growth (exponential catastrophe) appears. This exponential
 catastrophe is a main difficulty of accepting this complex
 eigenstate as a representation of unstable states.

 To see this problem let us consider the time evolution of
 $|\phi_1\ket$ for the atom component and field component.
 The atom component of $|\phi_1\ket$ is given by
 \beqa
  \bra 1|e^{-iHt} |\phi_1\ket = N_1^{1/2} e^{-iz_1 t}. \EQN{20}
 \eeqa
 \Ep{20} holds for all $t$. For $t<0$, the RHS of \Ep{20} grows exponentially. If we had chosen the $-$ branch of the Friedrichs eigenstates as a starting point, then we would have a similar problem: the states would decay for $t<0$ and would grow exponentially for $t>0$.

\begin{figure}[htb] % Imported eps example.
\begin{center}
\psfig{file=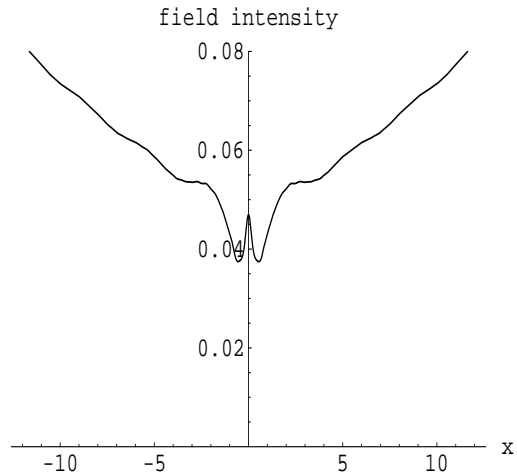, height=2.5in, width=3in}
 %unstable_fig1.ps

 \caption{ $|\bra \psi (x)|e^{-iHt}|\phi_{1}\ket|^2$ plot at
$t=10$. We see the exponential catastrophe for $|x|\rightarrow
\infty$. } \label{phi1}
\end{center}
\end{figure}

  This exponential growth also  appears in the field component of
  $|\phi_1\ket$. The  time evolution of this component  is becomes
 \beqa
 \bra \psi (x) | e^{-i H t} | \phi_1\ket  &=& e^{-iz_1 t} \bra \psi
 (x) | \phi_1 \ket \sim O(e^{-i z_1 (t-|x|)})\nonumber\\
& &\mbox{for large $|t-|x||.$}  \EQN{22}
 \eeqa
The field component shows exponential growth in $x$. The
 $|\bra \psi (x)|e^{-i H t}|\phi_1 \ket|^2 $ plot in $x$
 space for a fixed time is shown in Fig~\ref{phi1}.

We can avoid the exponential
growth by choosing suitable test function spaces. One such
approach is due to Bohm and Gadella \cite{Bohm}. They defined  complex states (Gamow
vectors) through the poles of the S-matrix, and restricted their
test function space to Hardy class functions from below. In Appendix A we review their approach and also discuss difficulties in
their approach.

In the next section we propose a new way to construct the unstable state
 for the Friedrichs  model. We separate the pole according to
the type of test function, and discuss the advantages of our
construction over the previous constructions of unstable states.

%%%%%%%%%%%%%%%%%%%%%%%%
  \section{A new unstable state in the Friedrichs  model}
 \label{new unstable}
%%%%%%%%%%%%%%%%%%%%%%%%
 In this section we propose a new way of defining the unstable state.
We want our unstable state to be memoryless and have no unphysical
growth. To this end, we construct a complex eigenstate of the time
evolution operator $e^{-iHt}$ which gives exponential decay, and a
suitable test function space which removes unphysical growth.

%%%%%
\subsection{Complex pole and integration contour}
%%%%%
 The complex eigenstate is related to the complex pole of Green's
 function $[\eta^+(z)]^{-1}$ (or the pole of S-matrix
 $\eta^-(\ome)/\eta^+(\ome)$)
 that can be calculated by perturbation from the original
 unperturbed eigenstate.  Consider the emission of the field by the excited
 atom. We focus on the overlap $\bra f |e^{-i H t}| 1\ket$ between the
 emitted field and a wave packet $\bra f|$. We restrict our attention to
 the case in which the wave packet $\bra f|x\ket$ is square integrable and
 localized, with compact support in space representation.

  From the completeness relation of eigenstates of $H$, we have
 \beqa
\bra f |e^{-i H t}|1\ket &=&  \bra f |e^{-i H t}\int_0^{\infty} d\ome\;
| F_\ome^+\ket \bra F_\ome^+| 1\ket    \EQN{59a}\\
 &=& \int_0^{\infty}d\ome e^{-i\ome
 t}\frac{\eta^-(\ome)}{\eta^+(\ome)}
 \bra f|F_\ome^-\ket\bra F_\ome^+|1\ket \
 \eeqa
  In \Ep{59a}, $1/\eta^+(\ome)$ has the pole $z_1 = \omet_1-i \gam=
 \ome_1+O(\lam^2)$ in the lower half plane. This is the pole we want to extract.

Our task is to take the residue at the pole $z_1$ in an integral of the form
 $\int_0^{\infty} d\ome\; h(\ome)$,  where $h(\ome)$ has the pole $z_1$
in the lower half of the complex plane. One simple way to do this is making a contour
 which encloses the pole and using the residue theorem.
 Suppose that we make a counterclockwise contour $C$ around the pole
 $z_1$. If $h(\ome)$ is analytic in $C$, we
  have
 \beqa
 \int_C d\ome\; h(\ome) = \int_C d\ome\;
 \frac{1}{\ome-z_1} h(\ome)(\ome-z_1) =
  2\pi i h_1(z) \nonumber\\
 \EQN{56}
 \eeqa
 where $h_1(\ome)$ is defined as
 \beqa
 h_1(\ome) \equiv (\ome-z_1)h(\ome). \EQN{56-0}
 \eeqa
 Note that $h_1(\ome)$ is analytic function inside $C$.

 The enclosing contour should be chosen according to the
 test function $\bra f|$. It would not be a good choice of contour
 if the test function blows up at the contour.
 Also, the test function space should be
 determined by considering underlying physics.

In scattering experiment usually a localized wave packet is
prepared.  Say $\bra f|x\ket$
 is zero outside the region $-x_0<x<x_0$.
 In our Hamiltonian system,  $\bra f| \ome \ket$ is given by
  \beqa
   \bra f|\ome\ket = \int_{-\infty}^{\infty} dx\; \bra f|x\ket
   \bra x|\ome\ket
   =\int_{-x_0}^{x_0} dx\; \bra f|x\ket
   \frac{ \cos (\ome x)}{\sqrt{\pi}}. \EQN{57}
   \eeqa
According to  a theorem due to Paley and Wiener
\cite{Boas}, the function   $\bra f | \ome\ket$  is entire
 function of exponential type $x_0$ and belongs to $L^2$ on the real axis
 of $\ome$ (see Appendix \ref{Paley}).  This theorem  shows that even though  $\bra f|\ome\ket$
 is $L^2$ on the positive real axis, it can be extended to the whole real axis and
remain in $L^2$. So, we can use the whole real axis as a part of
 enclosing contour. We write
 \beqa
  \int_0^\infty =   \int_{-\infty}^\infty -   \int_{-\infty}^0
  \eeqa
  The last term is a ``background'' integral, which does not give any pole contribution.

  To enclose the pole in the lower half plane, we need another
  piece of contour besides the real axis. If the function vanishes
  at the lower infinite semicircle, then the integral over the real
  axis is the same as the integral over the closed contour consisting of the real axis and the infinite
  lower semicircle, which encloses the $z_1$ pole.  For the Cauchy integral
  $$ \frac{-1}{2\pi i}\int_{-\infty}^{\infty}d\ome\; \frac{h_1(\ome)}{\ome-z_1}$$
with $h_1(\ome)$ vanishing on the lower infinite semicircle, we can
separate the $z_1$ pole residue by subtracting the other pole
residues.
\beqa
 \frac{-1}{2\pi i}
 \int_{-\infty}^{\infty}d\ome\; \frac{h_1(\ome)}{\ome-z_1}
  -\sum_n {\rm Res}[\frac{h_1(\ome)}{\ome-z_1}]\bigg|_{p_n}
 =h_1(z_1) \EQN{56-1}
 \eeqa
 In \Ep{56-1}, $p_n$ are possible poles of $h(\ome)$ other than $z_1$ in the lower
 half plane.

The physical
 meaning of the test function  vanishing at the lower infinite semicircle is causality, as we discuss next.

 %%%%%
\subsection{Hardy-class test functions and causality}
%%%%%

In \Ep{59a} we want to separate the part of the integrand that vanishes at the lower infinite semicircle in $\ome$ plane.  This can be done through the decomposition into Hardy-class functions from below and from above, which we define now.

 A complex function $G(E)$ on the real
 line is a Hardy class function from above $H^2_+$ (below $H^2_-$) if

  (1) $G(E)$ is the boundary value of a function $G(\ome)$ of
  complex variable (complex energy) $\ome = E+ i\eta$ that is
  analytic in the half plane $\eta >0$ ($\eta <0$).

  (2) $$ \int_{-\infty}^{\infty} |G(E+i \eta)|^2 dE < \mbox{finite} $$
   for all $\eta$ with $0<\eta < \infty$ ($-\infty<\eta < 0$).

   The function $G(E)$ is called an $H^2_{\pm}$ class function.
 There is an interesting relation between Hardy class functions
 and $L^2$ functions.

  Any $L^2$ function can be uniquely expressed as the sum of
  a function in $H^2_-$ and a function in $H^2_+$. If a function
  $f(\ome)$ is in $L^2$ on the real line, we can write
  \beqa
  f(\ome) = f^+(\ome) + f^-(\ome) \EQN{35}
  \eeqa
  with
  \beqa
 && f^+(\ome) \equiv \frac{1}{2\pi i} \int_{-\infty}^{\infty} d\ome'
  \frac{f(\ome')}{\ome'-\ome-i\eps} \in H^2_+, \nonumber\\
&&   f^-(\ome) \equiv
  \frac{-1}{2\pi i} \int_{-\infty}^{\infty} d\ome'
  \frac{f(\ome')}{\ome'-\ome+i\eps} \in H^2_-. \EQN{36}
  \eeqa
 We can also write
 \beqa
 &&  f^+(\ome) = \frac{1}{\sqrt{2\pi}} \int_0^{\infty} dt\; \hat{f}
  (t) e^{i\ome t},\nonumber\\
  && f^-(\ome) = \frac{1}{\sqrt{2\pi}} \int_{-\infty}^{0} dt\; \hat{f}
  (t) e^{i\ome t} \EQN{37}
 \eeqa
 where
  \beqa
 \hat{f}(t) = \frac{1}{\sqrt{2\pi}} \int_{-\infty}^{\infty} d\ome
 f(\ome) e^{-i\ome t}. \EQN{38}
 \eeqa
  $f^{\pm}$ has the following properties.
 \beqa
   \frac{1}{2\pi i} \int_{-\infty}^{\infty} d\ome
   \frac{f^+(\ome)}{\ome-z} = \left\{ \begin{array}{ll} f^+(z)
   \;& \mbox{for $Im(z) >0$} \\
   0\;& \mbox{for $Im(z) <0$} \end{array} \right. \EQN{39} \\
 \frac{-1}{2\pi i} \int_{-\infty}^{\infty} d\ome
   \frac{f^-(\ome)}{\ome-z} = \left\{ \begin{array}{ll} 0
   \;& \mbox{for $Im(z) >0$} \\
   f^-(z)\;& \mbox{for $Im(z) <0$} \end{array} \right. \EQN{40}
   \eeqa
 The Fourier transform of $f^{\pm}(\ome)$ has the property
  \beqa
  \hat{f}^{+} (t) = 0\;\;\; \mbox{for $t<0$},\;\;\; \hat{f}^{-} (t) = 0\;\;\; \mbox{for $t>0$}
\EQN{41}
 \eeqa

The most important property for us is that a Hardy-class function from above
(below) vanishes at the upper (lower) infinite semicircle in $\ome$ plane.

From \Ep{59a} we get
  \beqa
 & &\bra f|e^{-iHt}|1\ket \EQN{poled3}\\
 &=&  \int_0^\infty d\ome\, e^{-i\ome t} \frac{\eta^-(\ome)}{\eta^+(\ome)}
 \left[\bra f|\ome\ket + \frac{\lam
  v_\ome}{\eta^-(\ome)}\bra f |1\ket \right.\nonumber\\
&+& \left.\frac{\lam
  v_\ome}{\eta^-(\ome)}\int_0^{\infty}d\ome' \frac{\lam v_{\ome'} \bra f|\ome'\ket
  }{\ome-\ome'-i\eps } \right]\frac{\lam
  v_\ome}{\eta^-(\ome)}.
 \eeqa
For $t\ge0$, the second and third terms in the bracket $[\,\,]$
always vanish at the lower infinite semicircle in $\ome$ plane, if
they are finite at the real line. The first term  vanishes at the
lower infinite semicircle if \beqa \label{Gcaus1}
 \left[e^{-i\ome t}  \bra f|\ome \ket\right]^+ = 0
\eeqa
To see the physical meaning of this condition we write (see  \Ep{57})
 \beqa
  & &e^{-i\ome t}  \bra f|\ome \ket = \int_{-\infty}^{\infty} dx \bra
  f|x\ket e^{-i\ome t} \frac{ \cos (\ome x)}{\sqrt{\pi}}\EQN{causality-1}\\
&=&
  \frac{1}{2 \sqrt{\pi}}\int_{-\infty}^{\infty}dx \bra f|x\ket
  (e^{-i\ome (t-|x|)}+e^{-i\ome(t+|x|)}). \nonumber
 \eeqa
 Hence
 \beqa
  &&  \left[e^{-i\ome t}  \bra f|\ome \ket\right]^+ \\
&=&  \frac{1}{2 \sqrt{\pi}}\int_{0}^{\infty}dx \left(\bra f|x+t\ket
   + \bra f|-x-t\ket\right) e^{i \ome x} \nonumber
  \eeqa
  This vanishes when $t>|x_0|$,   the time required for the emitted outgoing wave to have an overlap with the wave packet. This is  causality condition follows from the requirement that the integrand in \Ep{poled3} vanishes at the lower infinite semicircle in order to take the residue at the pole $z_1$.   One point to  note is that we considered the space and time together when we apply this condition of vanishing at the lower infinite semicircle.

%%%%%%%%%
\subsection{Residue at the pole and unstable state}
\label{ResidueC}
%%%%%%%%%%%%

Now we take the residue at the pole $z_1$ in \Ep{poled3}. When \Ep{Gcaus1} is satisfied we get
 \beqa
 && {\rm Res}\left[\bra f|e^{-iHt}|1\ket\right]_{z_1}
=  \bra f|e^{-iHt} |\phi_1\ket \bra \phit_1|1\ket  \EQN{poled4}\\
  &=& e^{-i z_1 t} \bra f |\phi_1\ket \bra \phit_1|1\ket
\,\, {\rm if \,}  \left[e^{-i\ome t}  \bra f|\ome \ket\right]^+ = 0 \nonumber
 \eeqa
 where $|\phi_1\ket$ is the complex eigenvector of Hamiltonian in \Ep{18} .
 To generalize this result, we define the space of test functions ${\cal E}_H$ as the set
 of functions $\bra f|e^{-i H t}|\ome\ket$ with $t\ge 0$ and $\bra f|x\ket$ being
 in $L^2$ and having compact support. This also implies that $\bra
 f|\ome\ket$ is $L^2$ and exponential type by Paley and Wiener
 theorem. Due to the form factor $v_\ome$ of our model (see
 \Ep{4}), $\bra f|F_\ome^{\pm}\ket$ is also exponentially bounded.
 For these  test-functions we introduce a decomposition
  into a  component which vanishes at the lower  infinite semicircle
   in $\ome$ plane and a non-vanishing component.
 \beqa
  f(\ome) = f^v(\ome)+f^{nv}(\ome) \EQN{57-1}
  \eeqa
Precisely speaking, we denote
  $f^{nv}(\ome)$ as the part whose maximum modulus grows
  exponentially as the function approaches the
  lower infinite semicircle.

  Next, we define our custom complex delta
 function $\delta_a(\ome-z_1)$ as
 \beqa
&& \int_{0}^{\infty} d\ome\; f(\ome) \delta_a (\ome-z_1)\EQN{59-0}\\
 &=& \int_{0}^{\infty} d\ome\; (f^v(\ome)+f^{nv}(\ome))\delta_a (\ome-z_1)
 \equiv f^v (z_1). \nonumber
  \eeqa
 This delta function is similar to the complex delta functions
 defined in \cite{Nakanishi}, except that it takes only the part of
  test functions which vanishes at the lower infinite semicircle.

 If we do not restrict the test functions, we get the complex
 spectral decomposition
  \beqa
   \int_0^{\infty}d\ome |F_\ome^+ \ket\bra F_\ome^+| =
   |\phi_1\ket\bra \phit_1| + \int_0^{\infty}d\ome |F_{\ome d}\ket
 \bra F_\ome^+| \EQN{Cspectal1}
 \eeqa
 where
 \beqa
 |F_{\ome d}^+\ket \equiv |\ome\ket + \frac{\lam
 v_\ome}{\eta^{+d}(\ome)} |1\ket + \frac{\lam v_\ome}{\eta^{+
 d}(\ome)}\int_0^{\infty}d\ome'\frac{\lam
 v_{\ome'}|\ome'\ket}{\ome-\ome'+i\eps} \nonumber \\
 \eeqa
 with
  \beqa
  \frac{1}{\eta^{+d}(\ome)} \equiv
  \frac{1}{\eta^+(\ome)}\frac{z_1-\ome}{z_1^+ -\ome}.
 \eeqa

 With delta function $\delta_a(\ome-z_1)$ which restricts
 the test functions to
the part which vanishes at the lower infinite semicircle,
 we devise another expression.
 Generally it is not easy to define $f(\ome)^{v}$ and $f(\ome)^{nv}$
 for function $f(\ome)$. But for the function $\bra f|\ome\ket \in \cal
 E_H$, we can define
 \beqa
  & &\bra f |\ome\ket^{v} \equiv \bra f|\ome\ket^-,\;\; \bra
  f|\ome\ket^{nv} \equiv \bra f|\ome\ket^+,  \\
  & &\bra f|F_\ome^+\ket^{v}  \EQN{def-v1} \\
  & &\equiv \bra f|\ome \ket^- +\frac{\lam v_\ome}{\eta^+(\ome)}\bra
  f|1\ket + \frac{\lam v_\ome}{\eta^+(\ome)}\int_0^{\infty}
  d\ome'\frac{\lam v_{\ome'} \bra f|\ome'\ket^-}{\ome-\ome'+i\eps}
  \nonumber \\
  & &+
   \frac{\lam v_\ome}{\eta^+(\ome)}\int_0^{\infty}d\ome'
   \frac{\lam v_{\ome'} \bra f|\ome'\ket^+}{\ome-\ome'-i\eps},
  \nonumber
 \\
  & &\bra f|F_\ome^+\ket^{nv} \equiv \frac{-2\pi i\lam^2
  v_\ome^2}{\eta^+(\ome)} \bra f|\ome\ket^+, \EQN{def-v2} \\
  & &\bra f|F_\ome^-\ket^{v} \\
  & &\equiv \bra f|\ome\ket^-
  +\frac{\lam v_\ome}{\eta^-(\ome)} \bra f|1\ket+ \frac{\lam
  v_\ome}{\eta^-(\ome)}\int_0^{\infty}d\ome' \frac{\lam
  v_{\ome'}\bra f|\ome'\ket}{\ome-\ome'-i\eps}, \nonumber \\
  & &\bra f|F_\ome^-\ket^{nv} \equiv \bra f|\ome\ket^+
  \EQN{def-v4}
  \eeqa
 Because of equations (\ref{def-v1})- (\ref{def-v4}), the definitions
 of $f^v$ and $f^{nv}$ are different from Hardy class
 functions.

With these definitions we now define the unstable state and its
dual as
 \beqa
 & &|\phi_{1a}\ket \equiv N_1^{1/2}\left( |1\ket +
 \int_0^{\infty} d\ome\; \frac{\lam v_\ome |\ome\ket}{z_1^a-\ome}
 \right) \EQN{65} \\
 & &\bra\phit_{1a}| \equiv N_1^{1/2}\left( \bra 1| +
 \int_0^{\infty} d\ome\; \frac{\lam v_\ome \bra
 \ome|}{z_1^a-\ome} \right). \EQN{66}
 \eeqa
 where
 \beqa
 \frac{1}{z_1^a-\ome} \equiv \frac{1}{z_1-\ome} -2\pi i
 \delta_a(\ome-z_1). \EQN{67}
 \eeqa
 This gives
 \beqa
 & &\bra f|\phi_{1a}\ket \equiv N_1^{1/2}\bigg( \bra f|1\ket
 \nonumber \\
 & &+ \int_0^{\infty} d\ome\; \frac{\lam v_\ome \bra
 f|\ome\ket^v}{z_1^+ -\ome} + \int_0^{\infty} d\ome\; \frac{\lam v_\ome \bra
 f|\ome\ket^{nv}}{z_1-\ome}\bigg) \EQN{63} \\
 & & \bra\phit_{1a}|g\ket \equiv N_1^{1/2}\bigg( \bra 1|g\ket
 \nonumber \\
 & &+\int_0^{\infty} d\ome\; \frac{\lam v_\ome \bra
 \ome|g\ket^v}{z_1^+-\ome} + \int_0^{\infty} d\ome\; \frac{\lam v_\ome \bra
 \ome|g\ket^{nv}}{z_1-\ome}\bigg). \EQN{64}
 \eeqa
 Note that
 if $\bra f|\ome\ket^{nv} =0$ then $\bra f|\phi_{1a}\ket=\bra f|\phi_1\ket$.

 This unstable state $|\phi_{1a}\ket$ becomes a complex eigenstate
 of the $e^{-iHt}$ for special kind of test functions. Using
(see Appendix \ref{apdx:delta2})
 \beqa
  \bra F^-_\ome|\phi_{1a}\ket =-2\pi i N_1^{1/2} \lam v_\ome
  \delta_a(\ome-z_1)
 \EQN{delta2}
  \eeqa
  and
  \beqa
  \eta^-(z_1) = -2\pi i \lam^2 v_{z_1}^2, \EQN{eta- z_1}
  \eeqa
we have a complex eigenvalue
 equation (for $t\ge0$)
 \beqa
 \bra f|e^{-iHt} |\phi_{1a}\ket = e^{-iz_1 t} \bra f|\phi_{1a}\ket
  \EQN{eigeneq}
 \eeqa
if  $\bra f|e^{-iHt}|\ome\ket $ is in $\cal E_H$ and vanishes at the lower infinite
 semicircle, i.e., if $\bra f|$ has compact support in space and
 $\left(e^{-i \ome t}\bra f |\ome\ket\right)^+=0$.

  When  $\left(e^{-i \ome t}\bra f |\ome\ket\right)^+ \neq 0$, we
  have
  \beqa
 & &\bra f|e^{-iHt} |\phi_{1a}\ket \nonumber \\
 & & = N_1^{1/2} e^{-iz_1 t} \bigg(
 \bra f|1\ket + \int_0^{\infty} d\ome \frac{\lam v_\ome \bra
 f|\ome\ket}{z_1^+ -\ome} \nonumber \\
 & &+ (-2\pi i \lam v_{z_1}) (e^{iz_1 t}
 [e^{-i\ome t} \bra f|\ome\ket^+]^-_{z_1} - \bra f|\ome\ket^+_{z_1})
 \bigg). \nonumber \\
 \EQN{noeigeneq}
  \eeqa
 As we see later in next section the exponentially growing part is removed
 in \Ep{noeigeneq}.

  \Ep{eigeneq} is clearly different from the usual eigenvalue equation.
   We make
  comments about this equation here. First, this equation is an eigenvalue
  equation in a restricted test function space. The test function
  restriction is made according to the physics of the system and
  causality condition. Second, it is an eigenvalue equation of the
  time evolution operator $e^{-iHt}$, rather than $H$.
 In following sections we will discuss the space-time behavior of
 this new unstable state and possible complex spectral
 representations.

 Note that here we have focused on the semi-group that gives decay for $t>0$. In a similar fashion,
 we can get results for the other semi-group with decay for $t<0$ starting with the
 $-$ branch of the Friedrichs eigenstates, and exchanging the roles of
  functions which vanish at the lower infinite semicircle and functions
  which vanish at the upper infinite semicircle.

 %%%%%%%%%%%%%%%%%%
\subsection{Time evolution of unstable state}
\label{e^-iHt of phia}
%%%%%%%%%%%%%%
We act the time evolution operator $e^{-iHt}$ on the state
$|\phi_{1a}\ket$.
 The atom component of time evolved ket
becomes
 \beqa
 & &\bra 1|e^{-iHt}|\phi_{1a}\ket = \bra 1 |e^{-iHt} \int_0^{\infty}
 d\ome |F^-_\ome\ket \bra F^-_\ome |\phi_{1a}\ket \nonumber \\
&=& \bra 1| \int_0^{\infty}
 d\ome\; e^{-i\ome t} |F^-_\ome\ket(-2\pi i N_1^{1/2} \lam v_\ome ) \delta_a (\ome -z_1) \nonumber\\
&=&
  N_1^{1/2} \Theta (t) e^{-iz_1t}.
 \EQN{71}
 \eeqa
 The atom component of time evolution of $|\phi_{1a}\ket$
 shows exact exponential decay for $t \ge 0$. This is a semi-group
 time evolution. Note also the exponential growth for the negative
 $t$ was removed.

Similarly, the field component of $|\phi_{1a}\ket$\ is (see
Appendix \ref{apdx:complex field})
  \beqa
  & &\bra \psi (x) |e^{-iHt}|\phi_{1a}\ket =\bra \psi (x) |e^{-iHt}\int_0^{\infty}
  d\ome |F^-_{\ome}\ket \bra F^-_{\ome}  |\phi_{1a}\ket \nonumber \\
  & &= -2\pi i \lam N_1^{1/2} \frac{\pi^{1/2}}{1+ z_1^2/M^2}e^{-iz_1 (t-
  |x|)} \Theta (t-|x|) \nonumber \\
 & &+  2 \lam N_1^{1/2} \lam \frac{z_1 }
 {(1+z_1^2/M^2)M} e^{-iz_1 t} e^{-M|x|}\Theta(t) \EQN{72}  \\
 & & -2 \lam
 N_1^{1/2}  \int_0^{\infty} d\ome'
 \frac{\cos (\ome'x)}{(1+\ome^{'2}/M^2)(z_1+\ome')}e^{-iz_1 t}\Theta(t) \nonumber
 \eeqa
 The first term in \Ep{72} comes from the complex pole at $z_1$.
 This is the travelling field with complex
 frequency inside the light cone. It corresponds to the decay product. The second term and third term do not travel but
decay with time. The second term is due to the non-locality of the
interaction,  caused by the ultraviolet cutoff in \Ep{4}. The
third term describes the cloud surrounding the atom
\cite{POP}. It is due to the background integral \cite{Bohm2}.

\begin{figure}[htb] % Imported eps example.
\begin{center}
\psfig{file=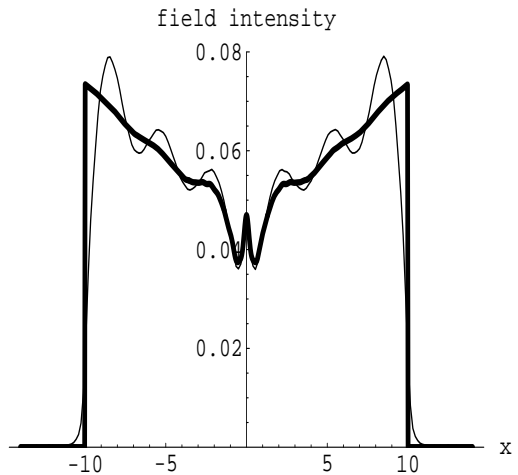, height=2.5in, width=3in}
 %unstable_fig2.ps
 \caption{
$|\bra \psi (x)|e^{-iHt}|\phi_{1a}\ket|^2$ plot (thick line) and
$|\bra \psi (x)|e^{-iHt}|1\ket|^2$ (thin line) at $t=10$. We see
the field component of new unstable state has a sharp front at
$|x|=t$. } \label{phi1a}
\end{center}
\end{figure}

 None of the terms in \Ep{72} has exponential blowup.
 The plot of $|\bra \psi (x) |e^{-iHt}|\phi_{1a}\ket|^2$ and
 $|\bra \psi (x) |e^{-iHt}|1\ket|^2$ in space
 is shown in Fig \ref{phi1a}. For weak coupling the field component of new unstable state is
 very close to the field component of bare atom decay.
 The field component of the new unstable state shows a sharp wave front, as
 the second and third terms in \Ep{26} give negligible contributions.
 We note that if  we had included virtual transitions in the Hamiltonian,
 the background contribution  would also be strictly confined within the light cone \cite{Comp}.

%%%%%%%%%%%%%%%%%%%%%%%%%%%
\subsection{Complex spectral representation of $\exp(-i H t)$}
%%%%%%%%%%%%%%%%%%%%%%%%%%%

Let us apply the complex delta function $\delta_a(\ome-z_1)$ to the complete set of  Friedrichs
eigenstates of $H$.The effect of pole
  enclosing contour is obtained by multiplying $(-2\pi i)(\ome-z_1)\delta_a(\ome-z_1)$
  to the Friedrichs solution and integrating over $\ome$. The factor
  $-2\pi i$ appears since the real axis and lower
  infinite semicircle clockwisely enclose the lower half plane
  pole.
    By this operation we get the $z_1$ residue of the part of the  test function
that vanishes at the lower infinite semicircle.

 Consider the inner product $\bra f| e^{-i H t}|g\ket$ where $\bra
 f|\ome\ket$ and $\bra \ome |g\ket$ are in $\cal E_H$.
 We have
 \beqa
  & &\bra f|e^{-i H t}|g\ket =\int_0^{\infty}d\ome\;  e^{- i\ome t} \bra f|F^+_\ome\ket \bra
  F^+_\ome |g\ket \nonumber \\
 &=&\int_{0}^{\infty} d\ome
   \; e^{- i\ome t} \bra f|F^+_\ome \ket \bra F^+_\ome|g\ket(-2\pi i)(\ome-z_1)\delta_a
   (\ome-z_1)
   \nonumber \\
  &+&  \int_{0}^{\infty} d\ome \; e^{- i\ome t}  \bra f|F^+_\ome \ket \bra
   F^+_\ome|g\ket \nonumber\\
  &\times& (1- (-2 \pi i)(\ome-z_1)\delta_a (\ome-z_1) ). \EQN{61}
 \eeqa
In \Ep{61}, the first term in the right hand side will give the
 pole separation we wanted.
By definition,
 $\delta_a(\ome-z_1)$ selects the part of $e^{-i\ome t}\bra f|F_\ome^-\ket
 \bra F_\ome^+|g\ket$ which vanishes at the
 lower infinite semicircle. That contains the term $e^{-i\ome t}\bra
 f|\ome\ket\bra \ome|g\ket$. Generally
 \beqa
  [e^{-i\ome t}\bra f|\ome\ket\bra \ome| g\ket]^- \neq e^{-i\ome
  t}\bra f|\ome\ket^- \bra \ome|g\ket^-,
  \eeqa
so the first term in RHS of \Ep{61} is not
 factorizable to the left and right complex eigenekts.
 But there are cases in which the factorization is possible,
 and we show two simple cases.

One case is when all terms vanish at the lower infinite
 semicircle. This condition is satisfied
when $t\ge 0$ and  the terms $[e^{-i\ome t} \bra \ome|g\ket]$,
$[e^{-i\ome
 t}\bra f|\ome\ket]$, and $[e^{-i\ome t}\bra f|\ome\ket\bra
 \ome|g\ket]$ vanish at the lower infinite semicircle (their $+$ components vanish).
The physical meaning of these conditions are the following.
Suppose that $\bra f|x\ket$ has a compact support $[-x_f, x_f]$ in
$x$ space. By \Ep{57} and Paley-Wiener theorem, $\bra f|\ome\ket$
is an $L^2$ function of exponential type $x_f$. Similarly for
$\bra x|g\ket$ with a compact support $[-x_g, x_g]$, $\bra
\ome|g\ket$ is an $L^2$ function of exponential type $x_g$. When
the functions approach to the lower infinite semicircle,
 $\bra f|\ome\ket = O(e^{i\ome x_f})$ and $\bra \ome|g\ket =
 O(e^{i\ome x_g})$. So
 $[e^{-i\ome t} \bra f|\ome \ket]^+=0$ implies that for time
 $t> x_f$ all the field component of $\bra f|$ is inside the
 causal region from the atom. Similarly, $[e^{-i\ome t} \bra \ome|g \ket]^+=0$
 means that for $t> x_g$ all the field component of $|g\ket$ is
 inside the causal region from the atom. The condition
$[e^{-i\ome t}\bra f|\ome\ket\bra \ome|g\ket]^+=0$ means that for
time $t>x_f+x_g$ a photon from $g$ can be scattered through the
atom and go to the field in $f$. This is the causality condition
of scattering of the field from the region occupied by $|g\ket$ to
 the region occupied by $\bra f|$.

In this case we get
    \beqa
  \bra f|e^{-iH t}|g\ket &=& e^{-iz_1 t} \bra f|\phi_{1}\ket\bra
 \phit_{1}|g\ket \nonumber\\
 &+& \int_0^{\infty}d\ome\;e^{-i\ome t} \bra f |F_{\ome d }^+\ket\bra
 F_\ome^+| g\ket. \EQN{spectral e^{-iHt}}
 \eeqa

 Another case that we can factorize the pole part in \Ep{61} is when
 $\bra f|\ome\ket=0$ or $\bra \ome|g\ket=0$. In this case (we consider
 $\bra \ome|g\ket=0$. Similar result can be shown for $\bra f|\ome\ket=0$.),
   we get
\beqa
  & &\bra f|e^{-iHt}|g\ket \nonumber \\
  & &= N_1^{1/2}e^{-iz_1 t} \bigg( \bra f|1\ket
  + \int_0^{\infty}d\ome \frac{ \lam v_\ome}{z_1^+-\ome} \nonumber
  \\
  & &+ (-2\pi i \lam v_{z_1}) ( e^{iz_1 t} [e^{-i\ome t}\bra
  f|\ome\ket^+]^-_{z_1} - \bra f|\ome\ket^+_{z_1})\bigg)\bra
  \phit_1|g\ket \nonumber \\
  & &+ \int_0^{\infty}d\ome\;e^{-i\ome t}\bra f|F_{\ome a}\ket\bra
  F_\ome^+|g\ket \EQN{fac-case2}
\eeqa
 like \Ep{noeigeneq}, with
 \beqa
  |F_{\ome a}^+\ket = |\ome\ket + \frac{\lam
  v_\ome}{\eta^{+a}(\ome)} |1\ket + \frac{\lam
  v_\ome}{\eta^{+a}(\ome)}\int_0^{\infty}d\ome'\frac{\lam
  v_{\ome'}|\ome'\ket}{\ome-\ome'+i\eps} \nonumber \\
  \eeqa
  where
  \beqa
  \frac{1}{\eta^{+a}(\ome)} \equiv \frac{z_1-\ome}{z_1^a -\ome}.
  \eeqa
\Ep{fac-case2} becomes \Ep{spectral e^{-iHt}} when
 \beqa
[e^{-i\ome t}\bra f|\ome\ket]^- = e^{-iz_1 t}\bra
f|\ome\ket^-_{z_1}. \EQN{fac-cond2}
 \eeqa
 As explained above, this is the
causality condition between the atom and the field. When
\Ep{fac-cond2} is not satisfied, we have \Ep{fac-case2} which does
not have an exponentially growing part outside the causal region.

%%%%%%%%%%%%%%%%%%%%%
 \section{Concluding remarks}
%%%%%%%%%%%%%%%%%%%%
 In this article we constructed a complex eigenstate with a suitable test
 functions
 which does
 not give an exponential catastrophe in the Friedrichs model. We extended the energy spectrum
 to the complex plane and separated the complex pole. To separate
 the pole, we choose a suitable contour  enclosing the pole.
 We established a class of test-functions (vanishing in the lower energy
 infinite semicircle) for which this pole separation is physically meaningful,
  giving a causal description of absorption and emission of decay products.

 The complex eigenstate constructed by separating
 the pole contribution in this way showed unique properties.
  Its atom component has exact
 exponential decay for the positive time. Its field component
consists of a  travelling wave with complex frequency inside the light
 cone. Thus the exponential catastrophe problem was removed.

  This complex eigenstate is an eigenstate of time evolution operator
  $e^{-iHt}$. The test function restriction is
  done by considering both time and space, rather than considering only time
  independent picture. In this way this
  complex eigenstate captures essential features of unstable
  particles in the physically meaningful region.

  In our opinion, this work is one nice example of
  constructing quantum mechanics outside the Hilbert space. The
  Hilbert space is very useful for describing stationary states.
  The eigenvalues are real, and the conserved norm represents the
  probability. But in our Hamiltonian, the atom state decays
  into the field, and the field is absorbed by the atom state.
   When we consider only
   the field space, norm is not conserved since fields are
   absorbed by the atom states, or the atom emits fields.
   In this case, we don't need to stick to the Hilbert space
   formalism, and distributions and suitable test functions
 can be used.

In this paper the specification and decomposition of our test
function space was done for the Friedrichs model, with the
dispersion relation $\ome_k=|k|$. Different test function space
should be used when the dispersion relation is different.
Moreover, we limited the initial test-functions
 to functions exponentially bounded at infinity. Inclusion of other functions
  such as Gaussians requires further consideration.
Also, we have limited our discussion to Dirac bras or kets. An extension to density operators in Liouville
space involves products of distributions, which will be considered
in future works.

\acknowledgements
  The authors thank Dr. T. Petrosky for helpful
  comments. One of authors thank Asia Pacific Center for Theoretical physics for their
  hospitality. We acknowledge the Engineering Research Program of the Office of Basic
Energy Sciences at  the U.S. Department of Energy, Grant No
DE-FG03-94ER14465, the Robert A. Welch Foundation Grant F-0365,
APCTP center for supporting  this work.

 \appendix

%%%%%%%%%%%%
%%%%%%%%%%%%%
 \section{Gamow vector with Hardy class  test functions}
%%%%%%%%%%%%%

 In this section we review the Gamow vector formalism introduced by Bohm and
 Gadella, and apply their formalism to the Friedrichs model.
 We  show that the Gamow vector obtained also has difficulties to represent
 the decaying state. The original  derivation of  Gamow vectors presented  in this
 section is found in  Bohm's book \cite{Bohm2}.

 In Ref. \cite{Bohm2},  Gamow vectors  are derived by considering S-matrix
 elements for the scattering of a pure state $\phi^{in}$ into a
 pure physical state $\psi^{out}$. $\phi^{in}$ is a controlled free state
 and determined by the preparation apparatus.
 $\psi^{out}$ is a free state controlled by the registration apparatus.

  When the Hamiltonian can be written as $H=H_0 +V$, where $H_0$ is
  the free Hamiltonian and $V$ is the interaction, the exact
  states $\phi^+ (t)$ and $\psi^-(t)$ are written as
 \beqa
 & &\phi^+ (t) = \phi^{in} (t) + \int_{-\infty}^{\infty} dt' \; G^+
 (t-t') V \phi^{in} (t'), \EQN{23} \\
 & &\psi^- (t) = \psi^{out} (t) + \int_{-\infty}^{\infty} dt'\;
 G^-(t-t') V\psi^{out} (t'). \EQN{24}
 \eeqa
 In \Ep{24}, the Green's function $G^{\pm}$ is given by
  \beqa
  & &G^+(t) = \left\{ \begin{array}{ll} 0\;\;\; \mbox{if $t<0$,} \\
             -i e^{-i H t} \;\;\; \mbox{if $t>0$,} \end{array}
             \right. \EQN{25} \\
  & &G^-(t) = \left\{ \begin{array}{ll} i e^{-i H t} \;\;\; \mbox{if $t<0$,} \\
             0 \;\;\; \mbox{if $t>0$.}  \end{array}
             \right. \EQN{26}
 \eeqa
Defining the M\o ller wave operators as
 \beqa
 \Omega^{\pm} \equiv I + \int_{-\infty}^{\infty} dt'\; G^{\pm}
 (t-t') V e^{-i H_0 (t'-t)}, \EQN{27}
 \eeqa
  \Ep{23} and \Ep{24} can be written as
  \beqa
  \phi^+ (t) = \Omega^+ \phi^{in} (t),\;\;\;\; \psi^{-} (t) =
  \Omega^- \psi^{out}(t) \EQN{28}
  \eeqa
 and the scattering operator $S$ is defined as
  \beqa
   S \equiv \Omega^{-\dagger}\Omega^+ .\EQN{29}
   \eeqa

 The S-matrix element for $\phi^{in}$ and $\psi^{out}$ becomes
 \beqa
&&  (\psi^{out} (t), S \phi^{in}(t) ) = (\Omega^- \psi^{out}(t)
  ,\Omega^+ \phi^{in} (t) ) \nonumber\\
&=& (\psi^-(t), \phi^+ (t)) = (\psi^-,
  \phi^+) \EQN{30}
  \eeqa
   We can calculate \Ep{30} using the eigenvectors of the total
   Hamiltonian $H$. If we write the eigenvectors of the free
   Hamiltonian $H_0$ as $|E\ket$, the eigenvectors of the total
   Hamiltonian $H$ can be obtained using the M\o ller wave
   operators
   \beqa
     & &|E^{\pm}\ket = \Omega^{\pm} |E\ket, \EQN{31} \\
 & &H_0 |E\ket = E|E\ket ,\;\;\; H|E^{\pm}\ket = E|E^{\pm}\ket.
 \EQN{32}
  \eeqa
 The eigenkets $|E^+\ket$ and $|E^-\ket$ are related by
  \beqa
  |E^+\ket = |E^- \ket S(E). \EQN{33}
  \eeqa
 Using the eigenvectors of total Hamiltonian, \Ep{30} can be
 written as
 \beqa
 (\psi^-, \phi^+) = \int_0^{\infty} dE \; \bra \psi^-|E^-\ket S(E)
 \bra E^+ |\phi^+\ket
  \EQN{34}
 \eeqa

We assume the S-matrix has a single complex pole $Z_R$,
\beqa
   S(E) = \frac{s_{-1}}{E-Z_R}+s_0+s_1 (E-Z_R) +... \EQN{43}
   \eeqa
 To introduce the Gamow vector associated with this pole,
 Bohm and Gadella defined a test function space $\Phi_-$ in which
 functions are Hardy class functions from below  and, in addition,
   are analytic functions that vanish
 faster than any inverse polynomial at the lower infinite
 semicircle.
 It is assumed that $\bra
 \phi^-|E^-\ket$ and $\bra E^+|\phi^+\ket$ both belong to
 $\Phi_-$.

   With these properties in mind, we continue our discussion about
  the Gamow vector. When $\bra \psi^- |E^-\ket$ and $\bra E^+
  |\phi^+\ket$ both belong to $H^2_-$, we have
  \beqa
  && (\psi^-, \phi^+) = \int_{-\infty}^{0} dE \bra  \psi^-
  |E^-\ket S(E) \bra E^+|\phi^+\ket \nonumber\\
&+& \int_{-\infty}^{\infty} \bra  \psi^-
  |E^-\ket \frac{s_{-1}}{E-Z_R} \bra E^+|\phi^+\ket. \EQN{42}
 \eeqa
 Here $s_{-1}$ is the residue of $S(E)$ at the complex pole $Z_R$.
    Using \Ep{40}, we can write
 \beqa
  && (\psi^-, \phi^+) = \int_{-\infty}^{0} dE \bra  \psi^-
  |E^-\ket S(E) \bra E^+|\phi^+\ket \nonumber\\
 &+& (-2 \pi i s_{-1}) \bra  \psi^-
  |Z_R^-\ket  \bra Z_R^+|\phi^+\ket. \EQN{44}
 \eeqa
 Thus omitting the arbitrary vector $\psi^-$ in $H^2_-$,
 \beqa
  && |\phi^+\ket = \int_0^{-\infty} dE |E^+\ket \bra E^+|\phi^+\ket \nonumber\\
&+&
  |Z_R^-\ket (-2\pi i s_{-1}) \bra Z_R^+ |\phi^+\ket. \EQN{45}
  \eeqa
 where the complex eigenvector $|Z_R^-\ket$ is given by
 \beqa
 |Z_R^-\ket = -\frac{1}{2\pi i}\int_{-\infty}^{\infty} dE
 |E^-\ket \frac{1}{E-Z_R}, \EQN{46}
 \eeqa
 which is a functional  over $H^2_+$ only.

If  $\bra
 \psi^-|E^-\ket$ belongs to $H^2_-$, then the time-evolved state $ \bra
 \psi^-|e^{-iHt}|E^-\ket$ becomes also a Hardy class function from below if
 $t \geq 0$. Using \Ep{40}, we obtain
 \beqa
&&  \bra e^{iHt} \psi^-|Z_R^-\ket = e^{-iZ_R t} \bra \psi^-|Z_R^-\ket \nonumber\\
&& \mbox{for $t \geq 0$ and every $\bra \psi^-|E^-\ket$ in
 $H^2_-$} \EQN{47}
 \eeqa

This Gamow vector does show exponential decay and semigroup time
evolution for $t\ge0$. To see if this definition is suitable for
the representation of unstable states, we should also check how the
field component of this Gamow vector behaves. We apply the above
Gamow vector formalism to the Friedrichs  model, and see how its
field component is
 represented in position space.

  In the Friedrichs model, the exact eigenvectors $|E^{\pm}\ket$
are explicitly written as $|F_\ome^{\pm}\ket$.  From the relation
 \beqa
 \eta^+(\ome)-\eta^-(\ome) = 2\pi i \lam^2 v_\ome^2, \EQN{48}
 \eeqa
 we can show that
 \beqa
  |F^+_\ome\ket = \frac{\eta^-(\ome)}{\eta^+(\ome)}
  |F^-_\ome\ket \EQN{49}
  \eeqa
 and the scattering matrix $S(\ome)$ is
 \beqa
  S(\ome) = \frac{\eta^-(\ome)}{\eta^+(\ome)}. \EQN{50}
  \eeqa
 The pole of $S(\ome)$ is at $z_1$, which satisfies $\eta^+(z_1)
 =0$. According to the above formalism, the Gamow vector in the Friedrichs model is
 \beqa
 |z_1^-\ket =- \frac{1}{2\pi i } \int_{-\infty}^{\infty}d\ome\;
 |F^-_\ome\ket \frac{1}{\ome-z_1}. \EQN{51}
 \eeqa

\begin{figure}[htb] % Imported eps example.
\begin{center}
\psfig{file=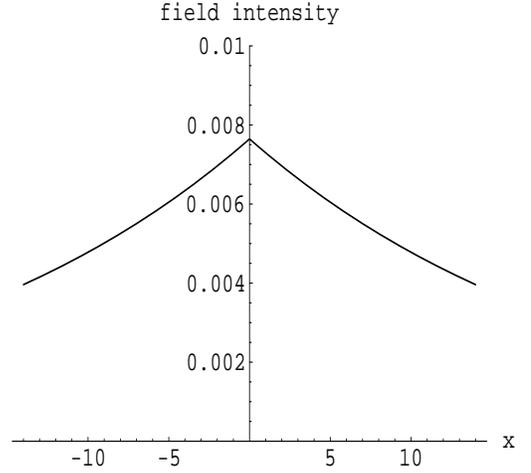, height=2.5in, width=3in}
 %unstable_fig2.ps
 \caption{
$|\bra \psi (x)|e^{-iHt}|z_{1}\ket|^2$ plot at $t=10$. It shows
the tail part of exponential field. } \label{bgplot}
\end{center}
\end{figure}

Let us calculate the field component for this Gamow vector. From
Eqs.  (\ref{21-x}) and (\ref{51}),
 \beqa
&&  \bra \psi(x) |F_\ome^-\ket = \sqrt{\frac{2}{\ome}} \cos (\ome
  x) \EQN{52} \\
&+& \frac{\lam v_\ome}{\eta^-(\ome)} \int_0^{\infty}d\ome'
  \sqrt{\frac{2}{\ome'}} \frac{\lam v_{\ome'} }{\ome-\ome'-i\eps}\cos (\ome'
  x). \nonumber
  \eeqa
 In \Ep{52}, we separate the component which is in $\Phi_-$.
  Using \Ep{36} on the second Riemann sheet
 to separate this component, we get
 \beqa
&&   ( \bra \psi (x)|F_{\ome _{II}}^-\ket )_{\Phi_-} = 2 \pi i
  \frac{\lam^2 }{\eta^-(\ome)}\sqrt{\frac{\ome}{2}}  e^{-i\ome |x|} \nonumber\\
&\times& \left(
  \frac{1}{4} \frac{1}{(1-i\ome/M)^2}
  +\frac{1}{8}\frac{1}{(1-i\ome/M)} \right) \EQN{53}
  \eeqa
 Substituting $\ome=z_1$ we obtain the Gamow vector
field component
 \beqa
&&  \bra \psi(x) |z_1^-\ket =2 \pi i
  \frac{\lam^2 }{\eta^-(z_1)}\sqrt{\frac{z_1}{2}}e^{-i z_1 |x|} \nonumber\\
&\times&  \left(
  \frac{1}{4} \frac{1}{(1-iz_1/M)^2}
  +\frac{1}{8}\frac{1}{(1-i z_1/M)} \right)  \EQN{54}
  \eeqa
 This shows exponentially decaying behavior for $x$. If we apply
 $e^{iHt} \psi(x)$ to the Gamow vector, we get
  \beqa
   & &\bra e^{iHt}\psi(x) |z_1\ket = e^{-i z_1 t} \bra \psi^-(x)
   |z_1\ket \nonumber \\
   &=& 2 \pi i
  \frac{\lam^2 }{\eta^-(z_1)}\sqrt{\frac{z_1}{2}} \left(
  \frac{1}{4} \frac{1}{(1-iz_1/M)^2}
  +\frac{1}{8}\frac{1}{(1-i z_1/M)} \right)  \nonumber\\
 &\times& e^{-iz_1(t+|x|)}   \mbox{for $t \geq 0$} \EQN{55}
\eeqa
 Since we restricted the test function
 space to the Hardy class   functions from below as well as  analytic functions
that vanish faster than any inverse polynomials at the lower
infinite semicircle (the space $\Phi_-$),
  the field component of this Gamow vector
 only gives the tail part of
 the exponential field $e^{-iz_1(t+|x|)}$, which does not show any
 wavefront (see figure  \ref{bgplot}).
 Actually, the dominant part of the field emitted from the decaying atom
 has the travelling wave with wavefront at the
 light cone, proportional to  $e^{-iz_1 (t-|x|)} \theta (t-|x|)$.
  The emited field term  $e^{-iz_1 (t-|x|)} \theta (t-|x|)$
 originates from the $e^{i\ome |x|}$
term of $\bra \psi(x) |F_\ome^-\ket$, which is outside the
 space $\Phi_-$.

 From the above example, we see that the emitted field is not
 fully represented by the Gamow vector $|z_1\ket$ with the test
 function $\bra \psi^-|E^-\ket \in \Phi_-$. We need another method to separate the complex pole from the
Hamiltonian  so that the separated pole term properly represents
the decaying atom as well as the emitted field.

%%%%%%%%%%%%%%%%%%
\section{Paley-Wiener theorem}
\label{Paley}
An entire
 function is one which is regular for all finite complex arguments.
 For the regular function $f(z)$ in $|z|<|R|$, we  denote $M(r)$ as the
 maximum modulus of $f(z)$ for $|z|=r<R$. For entire functions we take $R\to\infty$. The entire
 function $f(z)$ is called of positive order $\rho$ and of type
 $\tau$ if
  \beqa
  \lim_{r\rightarrow \infty} \sup r^{-\rho}\log M(r) =\tau. \;\;\;
  (0\le \tau\le \infty)
  \eeqa
 A function of order 1 and type $\tau$ ($\tau <\infty$) is called
 a function of exponential type.

 The theorem by Paley and Wiener states the following.
\newtheorem*{thma}{Theorem by Paley and Wiener}
\begin{thma}
An entire function $f(z)$ is of exponential type $x_0$ and
  belongs to $L^2$ on the real axis if and only if
  \beqa
  f(z) = \int_{-x_0}^{x_0} e^{iz x}\phi(x) \;dx, \EQN{58}
  \eeqa
  where
  \beqa
   \phi (x) \in L^2(-\infty, \infty).
   \eeqa
 Also, if $\phi(x)$ does not vanish almost everywhere in any
 neighborhood of $x_0$ (or $-x_0$) then $f(z)$ is order 1 and type
 $x_0$.
\end{thma}

\begin{widetext}
\section{derivation of \Ep{delta2}}
  \label{apdx:delta2}

 We have
 \beqa
& & \bra F^-_\ome |\phi_{1a}\ket \nonumber \\
& &=\bigg(\bra \ome |+\frac{\lam v_\ome}{\eta^+(\ome)}\bra 1|
+\frac{\lam v_\ome}{\eta^+(\ome)}\int_0^{\infty}d\ome' \frac{\lam
v_{\ome'}\bra \ome'|}{\ome-\ome'+i\eps}\bigg)N_1^{1/2}\bigg(|1\ket
+ \int_0^{\infty}d\ome\frac{\lam v_\ome}{z_1^a
-\ome}|\ome\ket\bigg) \nonumber \\
& &=N_1^{1/2} \bigg( \frac{\lam v_\ome}{z_1^a -\ome}+\frac{\lam
v_\ome}{\eta^+(\ome)} +\frac{\lam
v_\ome}{\eta^+(\ome)}\int_0^{\infty}d\ome' \frac{\lam^2
v_{\ome'}^2}{(\ome-\ome'+i\eps)(z_1^a-\ome')}\bigg) \nonumber \\
& &=N_1^{1/2} \bigg( \frac{\lam v_\ome}{z_1^a -\ome}+\frac{\lam
v_\ome}{\eta^+(\ome)} +\frac{\lam
v_\ome}{\eta^+(\ome)}\frac{1}{z_1-\ome} \bigg\{
\int_0^{\infty}d\ome' \frac{\lam^2v_{\ome'}^2}{\ome-\ome'+i\eps}
-\int_0^{\infty}d\ome'\frac{\lam^2v_{\ome'}^2}{z_1^+-\ome'}\bigg\}\bigg)
\nonumber \\
 & &=N_1^{1/2} \bigg( \frac{\lam v_\ome}{z_1^a
-\ome}+\frac{\lam v_\ome}{\eta^+(\ome)} +\frac{\lam
v_\ome}{\eta^+(\ome)}\frac{1}{z_1-\ome}
\bigg\{-\eta^+(\ome)+\ome-\ome_1-z_1+\ome_1\bigg\}\bigg) \nonumber
\\
& &=N_1^{1/2}\lam v_\ome \bigg(
\frac{1}{z_1^a-\ome}-\frac{1}{z_1-\ome}\bigg) = N_1^{1/2}\lam
v_\ome (-2\pi i)\delta_a(\ome-z_1).
 \eeqa

 \section{derivation of \Ep{72}}
 \label{apdx:complex field}
 We calculate the field component of $|\phi_{1a}\ket$.
 \beqa
 & &\bra \psi(x) |e^{-iHt}|\phi_{1a}\ket =
 \frac{1}{\sqrt{\pi}}\int_0^{\infty}d\ome' \frac{\cos (\ome'
 x)}{\sqrt{\ome'}}
 \bra\ome'
 |e^{-iHt}\int_0^{\infty}d\ome |F^-_\ome\ket\bra
 F_\ome^-|\phi_{1a}\ket \nonumber \\
 & &= \sqrt{2}\int_0^{\infty}d\ome e^{-i\ome t}\bigg[
 \frac{\cos (\ome x) }{\sqrt{\ome}} + \frac{\lam
 v_\ome}{\eta^-(\ome)} \int_0^{\infty}d\ome' \frac{\lam v_{\ome'}
 \cos (\ome' x)}{\sqrt{\ome'}(\ome-\ome'-i\eps)}\bigg)] \nonumber
 \\
 & &\times(-2\pi i) N_1^{1/2}\lam
 v_\ome \delta_a(\ome-z_1) \nonumber \\
 & &=-2\pi i
 N_1^{1/2}\int_0^{\infty}d\ome\bigg[\frac{2\lam
 e^{-i\ome t}\cos (\ome x)}{1+\ome^2/M^2}+ e^{-i\ome
 t}\frac{\lam^2 v_\ome^2
 }{\eta^-(\ome)}\int_0^{\infty}d\ome'\frac{2\lam \cos (\ome' x)}{(1+\ome^2/M^2)(\ome-\ome'-i\eps)}
 \bigg] \nonumber \\
 & &\times\delta_a(\ome-z_1). \EQN{fieldcal}
 \eeqa
 Although $\bra \psi (x)|\ome\ket$ has $1/\sqrt{\ome} $ singularity, combined with
 $v_\ome$ the test function for $\delta_a(\ome-z_1)$ in
 \Ep{fieldcal} becomes $L^2$ and analytic on the real line,
 grows at most exponentially at complex infinity with poles due to
 the form factor. Inside the square bracket of last term in \Ep{fieldcal},
  the separation of the part which vanishes at the
 lower infinite semicircle and non-vanishing part is clear
 due to the exponential functions. The vanishing part inside the
 square bracket in \Ep{fieldcal} is
\beqa
 \frac{\lam e^{-i\ome(t-|x|)}}{2 (1+\ome^2/M^2)}\Theta(t-|x|) +
 e^{-i\ome t}\Theta(t)\frac{\lam^2 v_\ome^2
 }{\eta^-(\ome)}\int_0^{\infty}d\ome'\frac{\lam \cos (\ome' x)}
 {(1+\ome^2/M^2)(\ome-\ome'-i\eps)}.
\eeqa
 Applying $\delta_a(\ome-z_1)$ to the above and rearranging terms,
 we get \Ep{72}.

\end{widetext}
%\bibliography{apssamp}% Produces the bibliography via BibTeX.

\end{document}